\documentclass[12pt]{article}

\usepackage{scicite}
\usepackage{psfrag}
\usepackage{graphicx}
\usepackage{amsmath}
\usepackage{amsfonts}
\usepackage{amssymb}
\usepackage{verbatim}

\usepackage{times}

\newenvironment{sciabstract}{%
\begin{quote} \bf}
{\end{quote}}

\newcounter{lastnote}
\newenvironment{scilastnote}{%
\setcounter{lastnote}{\value{enumiv}}%
\addtocounter{lastnote}{+1}%
\begin{list}%
{\arabic{lastnote}.}
{\setlength{\leftmargin}{.22in}}
{\setlength{\labelsep}{.5em}}}
{\end{list}}

\title{Sense Amplifier Comparator with Offset Correction
for Decision Feedback Equalization based Receivers}

\author
{Naveen Kadayinti,$^{\ast}$ and Dinesh Sharma\\
\\
\normalsize{Department of Electrical Engineering, Indian Institute of 
Technology Bombay,}\\
\normalsize{Powai, Mumbai 400076, India}\\
\\
\normalsize{$^\ast$E-mail: naveen@ee.iitb.ac.in.}
}

\date{}

\begin{document} 


\baselineskip24pt


\maketitle


\begin{sciabstract}

A decision feedback circuit with integrated offset compensation
is presented in this paper. The circuit is built around the sense amplifier
comparator. The feedback loop is closed around the first stage of the
comparator resulting in minimum loop latency. The feedback loop is
implemented using a switched capacitor network that picks from one of
pre-computed voltages to be fed back. The comparator's offset that is
to be compensated for, is added  in the same path. Hence, an extra offset
correction input is not required. The circuit
is used as a receiver for a 10~mm low swing interconnect implemented in
UMC 130 nm CMOS technology. The circuit is tested at a frequency of 1~GHz
and it consumes 145~$\mu$A from a 1.2V supply at this frequency.

\end{sciabstract}


\section{Introduction}
\label{sec:intro}
\begin{figure}[h!]
\centering
\psfrag{Data}{\small{Data}}
\psfrag{inp}{\small{$d_{in}$}}
\psfrag{tau}{$\tau$}
\psfrag{a}{$\alpha$}
\psfrag{Vof}{\small{$V_{of}$}}
\includegraphics[width=3.9cm]{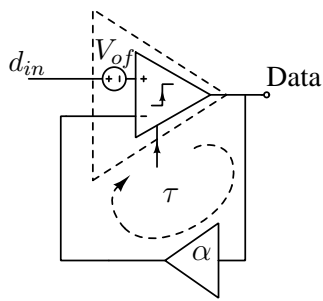}
\caption{Block diagram of a DFE circuit, indicating the sampler offset and the 
feedback loop.}
\label{fig:dfe_block}
\end{figure}

{Decision-Feedback-equalization} (DFE) is a technique 
that compensates for Inter-Symbol-Interference (ISI) in a serial input 
digital data 
signal~\cite{Mensink-jssc-2010,kim-jssc2010,Lee-jssc14,lookahead_dfe}. 
Fig.~\ref{fig:dfe_block} shows the block diagram of a DFE circuit. 
In this technique, a hard decision is made on the input in every clock cycle. 
This decision is scaled and subtracted from the input before the next 
sampling event. The scaling factor is chosen based on the amount of 
previous bit ISI in the input data. If the initial decision is correct, 
this effectively erases the memory of the previous bit.
The delay involved in 
making the hard decision, scaling it and subtracting from the next input
limits the maximum frequency of operation of this circuit. Further, in 
scaled technologies the decision devices have inherent offset that needs 
to be compensated for. In this paper, we propose a DFE circuit built around 
the Sense amplifier comparator that has low loop latency and features 
integrated offset compensation.

DFE is a simple technique and has found wide applications from low power to 
high performance communication systems. DFE has been proposed as an 
effective way of extending the bandwidth of repeaterless low swing 
interconnects~\cite{Mensink-jssc-2010,kim-jssc2010}. 
DFE has also been used to correct for errors in digital 
systems~\cite{dfe_critical_path}, for implementing low power
logic circuits based on pass transistor logic~\cite{pass_logic_dfe} and 
for enhancing bandwidth of flip-flops~\cite{mahendra_dfe}. A sense amplifier
comparator~\cite{Nikolic-jssc-2000} is used in most of these circuits as 
it can achieve high speed at low power consumption. When used for sampling 
low swing data, these comparators need offset compensation. Previous works 
have implemented this using an auxiliary input to the core 
comparator~\cite{Lee-jssc14,kim-jssc2010}. In high speed designs where the 
loop delay becomes the bottleneck, look-ahead-dfe is 
used~\cite{lookahead_dfe,kim-jssc2010}. 
In look-ahead-dfe, multiple comparators make decisions on the 
input data, each assuming a possible value of the previous decision.
This increases the number of comparators needed, each requiring its own offset 
compensation circuit as well.

In this paper, we propose a DFE circuit that has low latency and integrated
offset compensation. The feedback loop is built with a switched capacitor 
circuit, driven by the first stage of the sense amplifier, which picks from 
pre-computed inputs for the feedback. The offset to be corrected is added to the 
same feedback input, removing the need for an extra offset correction 
input to the comparator. The circuit is designed and fabricated 
in UMC 130 nm CMOS technology for a data rate of 1GHz. 
A double differential architecture, with a differential main input and 
differential feedback input, is used. For testing the 
equalizer, the comparator is used as a receiver of a 10~mm on-chip 
interconnect with a capacitively coupled low swing transmitter 
reported by Mensink et al. in~\cite{Mensink-jssc-2010}.  

The paper is organized as follows. The concept of switched capacitor DFE with 
offset compensation is discussed in Section~\ref{sec:concept}. The circuit 
implementation details are then discussed in Section~\ref{sec:implementation}, 
which is followed by results in Section~\ref{sec:results}. 
Section~\ref{sec:conclusion} then concludes the paper.

\section{DFE with switched capacitor feedback}
\label{sec:concept}
In time domain, the output $y[n]$ of the DFE circuit can be expressed 
in terms of the comparator input $x[n]$ as 
\begin{align}
y[n] &= Q\{x[n]\} = Q\{d_{in}[n] - \alpha y[n-1]\} \nonumber \\
	 &= \begin{cases} 
    -1, & \text{if $x[n]<0$}.\\
    +1, & \text{otherwise}.
  \end{cases}
\label{eqn:dfe}
\end{align}
Here, $y[n-1]$ is the hard decision made by the comparator in the previous 
cycle and $\alpha$ is a constant that is less than $1$. 
$\alpha$ is
chosen depending on the amount of ISI present in the input data.
The difference equation 
\begin{align*}
x[n] = d_{in}[n] - \alpha y[n-1],
\end{align*}
is a high pass function, 
which compensates for the ISI produced by the low pass nature of the 
interconnect.
Since $y[n]$ is a hard decision, the term 
$\alpha y[n-1]$ can take only one of two values, i.e.
\begin{align*}
\alpha y[n-1] &=\begin{cases}
	-\alpha, & \text{if $y[n-1]=-1$}. \\
	+\alpha, & \text{if $y[n-1]=+1$}. 
	\end{cases} 
\end{align*}

The analysis till now assumes
an ideal comparator. Practically, comparators also suffer from offset, 
which needs to be corrected. To compensate for the inherent offset of the 
comparator, the offset correction $V_{offset}$ can added within the same feedback i.e.
\begin{align*}
\alpha y[n-1] &=\begin{cases}
	-\alpha -V_{offset}, & \text{if $y[n-1]=-1$}.\\
	+\alpha -V_{offset}, & \text{if $y[n-1]=+1$}.
	\end{cases} 
\end{align*}

We implement the DFE circuit using a switched capacitor circuit that uses the 
comparator output to select from pre-computed voltages 
that correspond to $-\alpha -V_{offset}$ and $+\alpha -V_{offset}$ 
for the feedback.
Since most of the applications use differential input architecture, 
a comparator with a double differential input, i.e. 
with a differential main input and differential feedback input is used.
Such an implementation needs two precomputed differential bias inputs with 
different common modes, for the feedback network to pick from. 
Hence, a total of 4 distinct bias voltages are needed. This 
is explained in the following.

\begin{figure}[h!]
\centering
\psfrag{a}{\small{$\alpha $}}
\psfrag{D+}{\small{$V_i^-$}}
\psfrag{D-}{\small{$V_i^+$}}
\psfrag{Vfb+}{\small{$V_{fb}^+$}}
\psfrag{Vfb-}{\small{$V_{fb}^-$}}
\psfrag{VH1}{\scriptsize{$V_H^1$}}
\psfrag{VH2}{\scriptsize{$V_H^2$}}
\psfrag{VL1}{\scriptsize{$V_L^1$}}
\psfrag{VL2}{\scriptsize{$V_L^2$}}
\psfrag{-a}{\small{$\alpha$}}
\psfrag{Voffset}{\small{$V_{offset}$}}
\includegraphics[width=0.7\columnwidth]{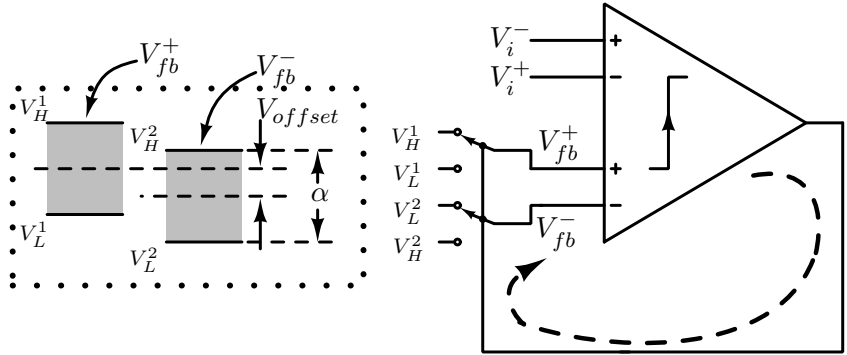}
\caption{Conceptual block diagram of the DFE circuit with
offset correction. The circuit has a differential main input and a differential 
feedback input.}
\label{fig:dfe_offset_concept}
\end{figure}

When $y[n-1]=+1$, the differential feedback voltages $V_{fb}^+,V_{fb}^-$ can 
be written as 
\begin{align*}
V_{fb}^+ &= V_H^1 = V_{cm} + \frac{V_{offset}}{2} + \frac{\alpha}{2}, \\	
V_{fb}^- &= V_L^2 = V_{cm} - \frac{V_{offset}}{2} - \frac{\alpha}{2}.
\end{align*}
Similarly, $y[n-1]=-1$,
\begin{align*}
V_{fb}^+ &= V_L^1 = V_{cm} + \frac{V_{offset}}{2} - \frac{\alpha}{2}, \\ 
V_{fb}^- &= V_H^2 = V_{cm} - \frac{V_{offset}}{2} + \frac{\alpha}{2}.
\end{align*}
Here, $V_{cm}$ is the common mode of the feedback input.
This is illustrated graphically in Fig.~\ref{fig:dfe_offset_concept}, 
along with the block diagram of the comparator with a double differential 
input. 

%
%

To summarize, the voltages $V_H^1,V_L^1,V_H^2$ and $V_L^2$ are the four feedback 
bias voltages. The difference $V_H^1-V_L^1$ (= $V_H^2-V_L^2$) 
corresponds to the feedback factor $\alpha$. The common modes of these two 
differential pairs are skewed by the offset to be corrected, as illustrated in 
Fig.~\ref{fig:dfe_offset_concept}.

\begin{figure}[h!]
\centering
\psfrag{Mp1}{\small{$M_p^1$}}
\psfrag{Mp2}{\small{$M_p^2$}}
\psfrag{Mp3}{\small{$M_p^3$}}
\psfrag{Mp4}{\small{$M_p^4$}}
\psfrag{Mp6}{\small{$M_p^6$}}
\psfrag{Mp7}{\small{$M_p^7$}}

\psfrag{Mn1}{\small{$M_n^1$}}
\psfrag{Mn2}{\small{$M_n^2$}}
\psfrag{Mn3}{\small{$M_n^3$}}
\psfrag{Mn4}{\small{$M_n^4$}}
\psfrag{Mn5}{\small{$M_n^5$}}
\psfrag{Mn6}{\small{$M_n^6$}}
\psfrag{Mn7}{\small{$M_n^7$}}

\psfrag{In+}{\small{$V_i^+$}}
\psfrag{In-}{\small{$V_i^-$}}
\psfrag{Voff+}{\small{$V_{fb}^+$}}
\psfrag{Voff-}{\small{$V_{fb}^-$}}
\psfrag{Ck}{\small{ck}}
\psfrag{Sb}{\small{$\overline{S}$}}
\psfrag{Rb}{\small{$\overline{R}$}}
\includegraphics[width=0.8\columnwidth]{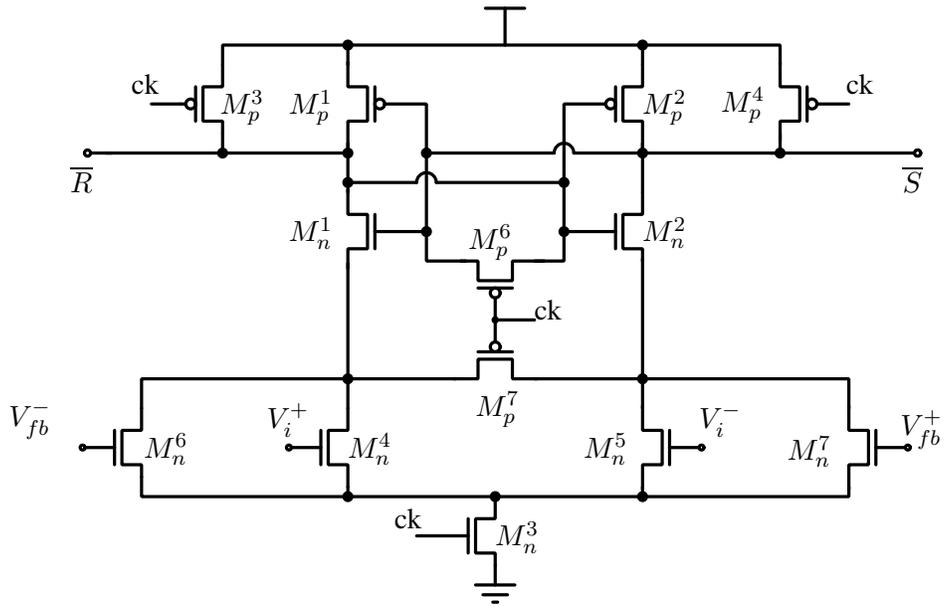}
\caption{First stage of the sense amplifier comparator, with additional input 
transistors for the feedback inputs.}
\label{fig:SAFF_stage1}
\end{figure}

We use the sense amplifier based comparator in the DFE circuit. The circuit 
diagram of the first stage of the comparator is shown in 
Fig.~\ref{fig:SAFF_stage1}. 
An additional input transistor pair is used for the feedback 
input~\cite{Lee-jssc14}.

\begin{figure}[h!]
\centering
\psfrag{time}{\small{\hspace{-1ex}Time (ns)}}
\psfrag{sbrb}{\small{$\overline{S}$, $\overline{R}$}}
\psfrag{VDD}{\hspace{-1ex}\scriptsize{$V_{DD}$}}
\psfrag{0}{\scriptsize{$0$}}
\psfrag{T}{\scriptsize{$T$}}
\psfrag{2T}{\scriptsize{$2T$}}
\psfrag{3T}{\scriptsize{$3T$}}
\includegraphics[width=0.65\columnwidth]{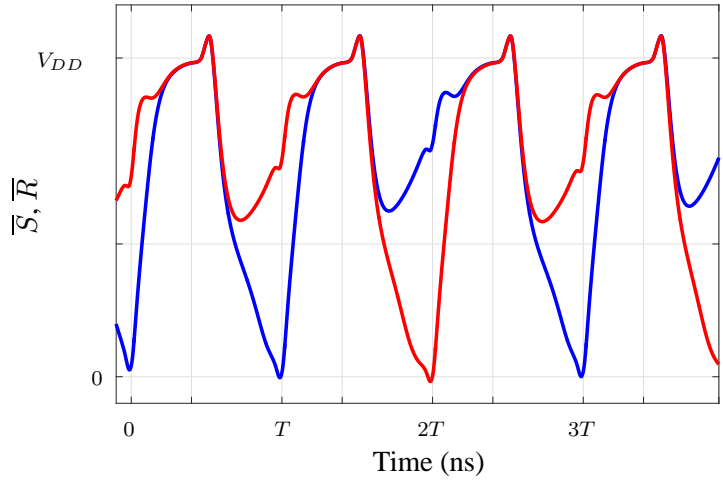}
\caption{$\overline{S}$ and $\overline{R}$ signals of the first stage of the
sense amplifier comparator}
\label{fig:sbrb}
\end{figure}

The second stage of the comparator is an SR slave latch~\cite{Nikolic-jssc-2000}.
Prior implementations of DFE using this comparator have used the slave latch 
output for the decision feedback~\cite{Mensink-jssc-2010}. We use the first 
stage output itself for the feedback, which results in minimum latency. 
The nodes $\overline{S}$ and 
$\overline{R}$ are precharged in every cycle to $V_{DD}$. During the input 
evaluation phase, these nodes discharge through the input transistors and 
depending on the input, one of the nodes discharges faster than the other.
When $\overline{S}$ and $\overline{R}$ are discharged below the trip point 
of the inverters formed by $M_p^1,~M_n^1$ and $M_p^2,~M_n^2$, the inverter 
positive feedback pulls $\overline{S}$ and $\overline{R}$ apart in the 
direction established by the input pair. Typical waveforms of 
$\overline{S}$ and $\overline{R}$ are shown in Fig.~\ref{fig:sbrb}.
The output of the first stage is used to drive a switched capacitor network 
which picks from the two pairs of differential bias voltages for the feedback. 
This effectively results 
in a low swing dynamic latch for the decision feedback. The circuit 
implementation is discussed in the next section.

\section{Circuit implementation}
\label{sec:implementation}
In this section we shall discuss the circuit implementation. The first 
subsection will describe the implementation of the comparator and the 
second subsection will describe the bias voltage generation. The circuit 
was designed in UMC 130 nm CMOS technology.

\subsection{Comparator with DFE}
\begin{figure}[h!]
\centering
\psfrag{Vfb+}{\small{$V_{fb}^+$}}
\psfrag{Vfb-}{\small{$V_{fb}^-$}}
\psfrag{VH1}{\small{$V_H^1$}}
\psfrag{VH2}{\small{$V_H^2$}}
\psfrag{VL1}{\small{$V_L^1$}}
\psfrag{VL2}{\small{$V_L^2$}}
\psfrag{Sb}{\small{$\overline{S}$}}
\psfrag{Rb}{\small{$\overline{R}$}}
\includegraphics[width=0.6\columnwidth]{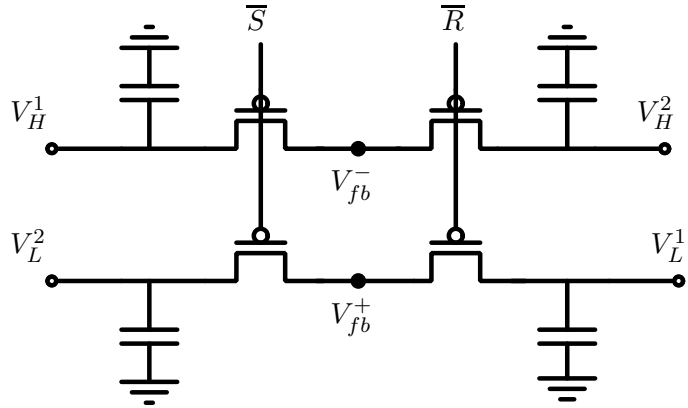}
\caption{Feedback network using $\overline{S}$ and $\overline{R}$ signals
as the select lines for an analog multiplexer. Low $V_t$ transistors are used
for the select switches.}
\label{fig:dfe_fb_nw}
\end{figure}

As discussed in the previous section, we shall use the sense amplifier 
comparator. The feedback network is a switched capacitor circuit driven 
by the first stage of the comparator, which is shown in Fig.~\ref{fig:dfe_fb_nw}.
In every clock cycle the comparator is reset, i.e. both $\overline{S}$ and 
$\overline{R}$ are 
precharged to $V_{DD}$. This puts the multiplexer in a high impedance
state. During input sample evaluation, one of $\overline{S}$ or $\overline{R}$ 
fall lower than the other, and the output of the analog multiplexer generates 
the scaled version of the resolved bit. This output is held dynamically on
the parasitic capacitance of the node, as the comparator precharges for 
the next cycle. Hence, the next cycle evaluation subtracts the scaled 
previous bit value.
%
Since the select transistors spend a little time in the ON state before 
the precharge phase of the next clock cycle begins, the time available for 
the output to change states is limited. Low $V_t$ transistors are used as 
switches in order to improve the selector performance.

\begin{figure}[h!]
\centering
\psfrag{In+}{\small{$V_i^+$}}
\psfrag{In-}{\small{$V_i^-$}}
\psfrag{Vfb+}{\small{$V_{fb}^+$}}
\psfrag{Vfb-}{\small{$V_{fb}^-$}}
\psfrag{Voff+}{\small{$V_{fb}^+$}}
\psfrag{Voff-}{\small{$V_{fb}^-$}}
\psfrag{Ck}{\small{ck}}
\psfrag{Sb}{\small{$\overline{S}$}}
\psfrag{Rb}{\small{$\overline{R}$}}
\psfrag{VCM}{\small{$V_{CM}$}}
\psfrag{= 0.6}{\small{Low $V_t$ transistors}}
\psfrag{wid}{\small{$W$~=~0.6~$\mu$m}}
\psfrag{VH1}{\small{$V_H^1$}}
\psfrag{VH2}{\small{$V_H^2$}}
\psfrag{VL1}{\small{$V_L^1$}}
\psfrag{VL2}{\small{$V_L^2$}}
\includegraphics[width=0.7\columnwidth]{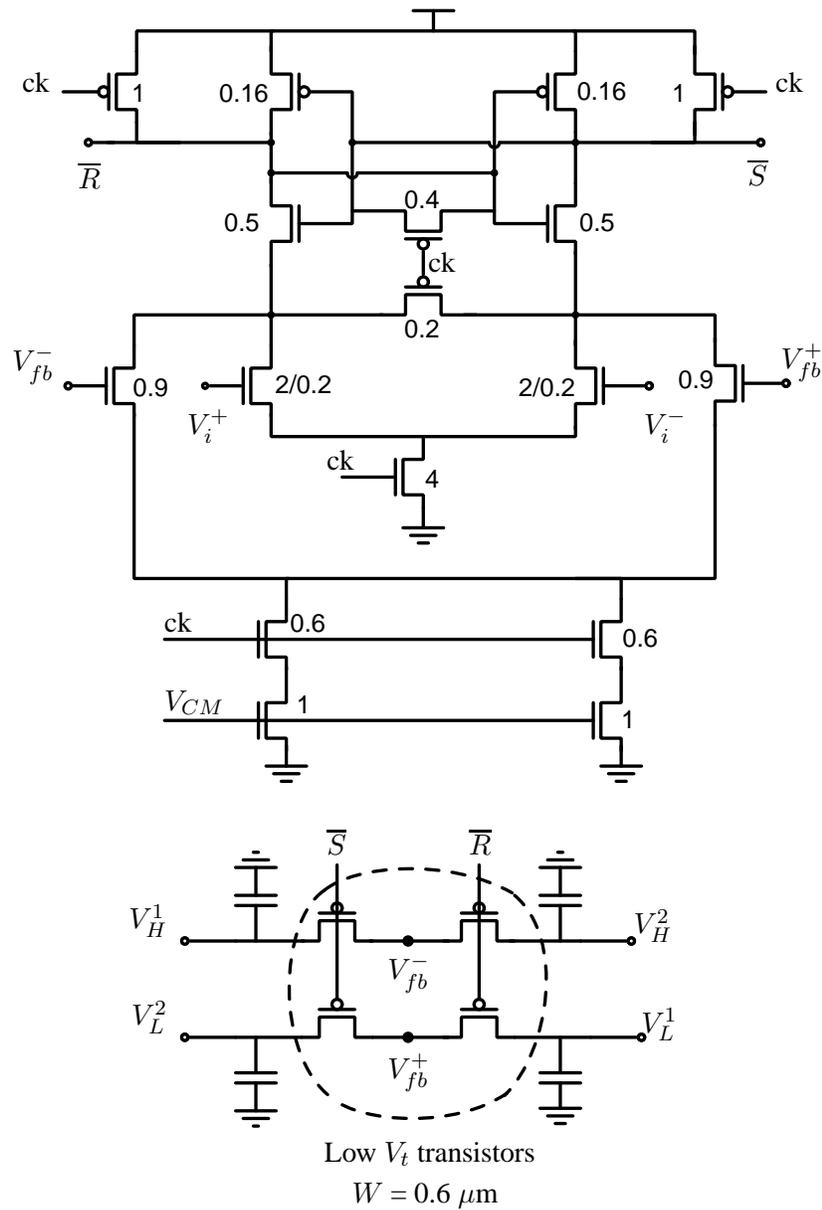}
\caption{DFE circuit implemented in UMC 130 nm technology, with transistor
dimensions. All dimensions are in $\mu$m. Unless explicitly mentioned, length
of all transistors is minimum which is 120~nm.}
\label{fig:dfe_dim}
\end{figure}

One of the difficulties of using $\overline{S}$ and $\overline{R}$ for driving 
the feedback comes from the very large common mode swings on these signals 
due to the pre-charge cycle. Hence, the feedback input needs to have a good 
common mode rejection ratio. The first stage of the sense amplifier comparator 
is modified to bias the feedback transistors with a tail current source.
The modified first stage is shown in Fig.~\ref{fig:dfe_dim}.
The dimensions of the transistors are also shown in Fig.~\ref{fig:dfe_dim},
where all dimensions are in~$\mu$m and unless specified otherwise, the length 
of the transistors is minimum which is 120~nm. 
The feedback voltages are chosen taking into account the gain of the 
feedback input of the comparator, relative to the main input pair's gain. 
In this design, the feedback network is 
designed to have half the gain of the main input pair.

\subsection{Bias generation}
\subsubsection{$V_{CM}$ generation}
\begin{figure}[h!]
\centering
\psfrag{Line+}{\small{$V_i^+$}}
\psfrag{Line-}{\small{$V_i^-$}}
\psfrag{Vcm}{\small{$V_{CM}$}}
\psfrag{Sb}{\small{$\overline{S}$}}
\psfrag{Rb}{\small{$\overline{R}$}}
\psfrag{ck}{\small{ck}}
\psfrag{Latch}{\small{Latch}}
\psfrag{Vinp}{\small{$V_{in}$}}
\psfrag{Vinn}{\small{$\overline{V_{in}}$}}
\includegraphics[width=0.4\columnwidth]{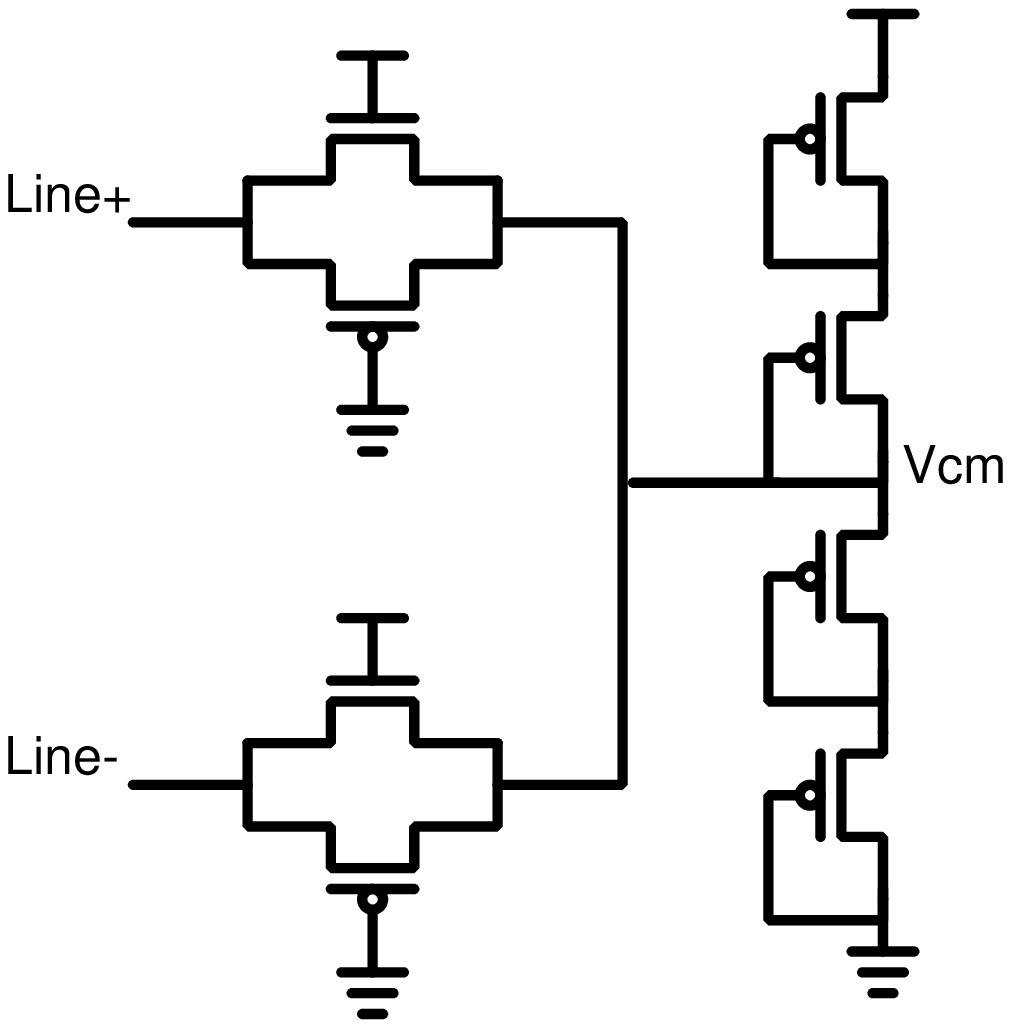}
\caption{Line termination circuit. The common mode is used to bias the 
feedback input tail transistor.}
\label{fig:rx_term}
\end{figure}

The feedback input pair needs a bias current source. This input is biased with 
the common mode of the main data input. In this way the relative strengths of 
the main and the feedback inputs track each other. This bias is derived from 
the receiver termination, which is shown in Fig.~\ref{fig:rx_term}. This 
is the same termination circuit reported in ~\cite{naveen_vlsi13}.

\subsubsection{Generation of feedback voltages}
\begin{figure}[h!]
\centering
\psfrag{A1}{\scriptsize{$OA_1$}}
\psfrag{VoCM}{\small{$V_{CM}$}}
\psfrag{binary tree}{\small{4 binary tree}}
\psfrag{Switch Matrix}{\small{switch matrices}}
\psfrag{Logic}{\small{Logic}}
\psfrag{alpha}{\small{$\beta$}}
\psfrag{offset}{\small{$REG_{of}$}}
\psfrag{1u}{\small{$I_{bias}$}}
\psfrag{VH1}{\small{$V_H^1$}}
\psfrag{VH2}{\small{$V_H^2$}}
\psfrag{VL1}{\small{$V_L^1$}}
\psfrag{VL2}{\small{$V_L^2$}}

\includegraphics[width=0.5\columnwidth,angle=0]{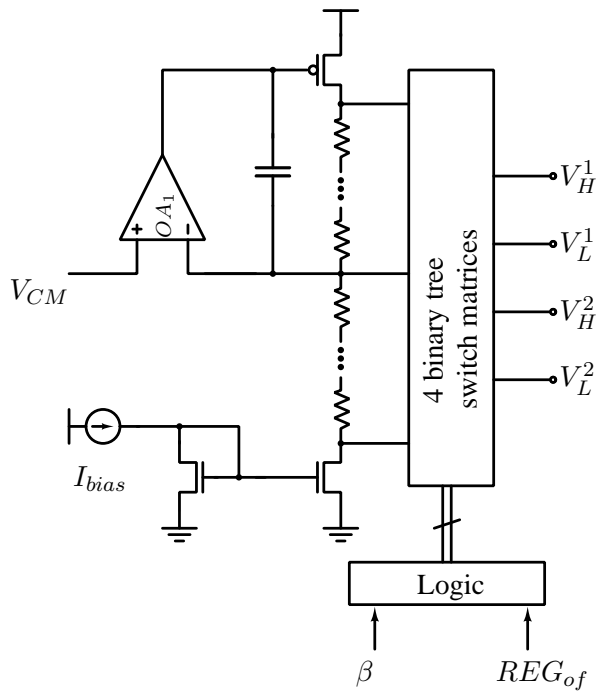}
\caption{This circuit generates multiple voltages and the
correct offset is chosen by the digital logic using the switch matrix.}
\label{fig:offset_gen}
\end{figure}

A 5 bit resistive string digital to analog converter (DAC) 
is used to generate the four bias voltages.
The resistor string, driven by a current source, is used to generate 32 levels
of voltages.
Fig.~\ref{fig:offset_gen} shows the circuit diagram of the resistor
string used to generate multiple bias voltages. The generated bias voltages are
centered around the common mode of the input of the comparator 
(which is forced by the negative feedback common mode feedback circuit 
built using the single stage opamp $OA_1$).
Four binary tree switch matrices, constructed using transmission gate switches, 
are used in the DAC to generate the required four bias outputs.
The outputs of the switch matrices are buffered with a single stage opamp.
Four digital words, each 5 bit wide, are used to select appropriate bias
voltages for the output. Two digital inputs, one for offset and another for
the feedback tap weight are used to generate the required 4 digital words
to drive the switch matrices. Of these, the offset control input is a 5 bit
control word and the feedback factor is a 4 bit control word. The required
4 digital words are calculated as
\begin{align*}
REG_{V_H^1} &= REG_{of} + \beta,\\
REG_{V_L^1} &= REG_{of} - \beta,\\
REG_{V_H^2} &= ``11111" - REG_{of} + \beta,\\
REG_{V_L^2} &= ``11111" - REG_{of} - \beta.
\end{align*}
Here, $REG_{of}$ is a 5 bit word that can take values from ``01000'' to ``10111'',
to select the offset input. $\beta = \alpha/2$, is a 4 bit 
word that can take values from
``0000'' to ``1000'' to chose the tap weight. This arrangement allows equal
dynamic range for the offset and the feedback tap weights. Depending on
expected offsets and desired tap weights, unequal splits can also be considered.

The DFE circuit was implemented using a resistive DAC to
generate the bias voltages. A 5 bit DAC with a dynamic range of $\pm$25~mV
was used. Non-salicided doped poly resistors were used for the DAC with a 
bias current of 5~$\mu$A. This allowed correction of offsets of up 
$\pm$12.5~mV and a feedback
input voltage of $\pm$12.5~mV. A 10~mm interconnect with a 1 tap capacitive
equalizer was used as the transmitter. The swing on the interconnect was
$\pm30$~mV. This implies that a feedback factor of up to $\sim$~0.2 is possible
with this implementation.

\section{Results}
\label{sec:results}
\begin{figure}[h!]
\centering
\psfrag{Receiver}{\scriptsize{Rx}}
\psfrag{Transmitter}{\hspace{1ex}\scriptsize{Tx}}
\psfrag{10mm}{\hspace{-3ex}\scriptsize{10 mm snaked interconnect}}
\includegraphics[width=4.5cm]{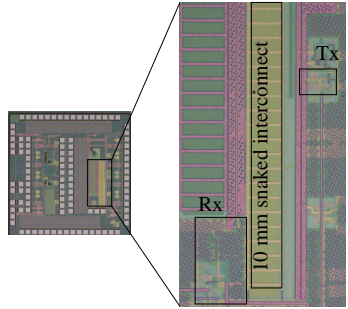}
\caption{Die photograph.}
\label{fig:ch3:dfe_chip}
\end{figure}

The circuit was designed and fabricated in UMC 130 nm CMOS technology. 
The total circuit area including the bias generation circuits is 
$91\mu$m$\times 52\mu$m. Fig.~\ref{fig:ch3:dfe_chip} shows a photograph of 
a bare die of the fabricated circuit.

\begin{figure}[h!]
\centering
\psfrag{ab = 0}{\scriptsize{$\alpha$ = 0}}
\psfrag{ab = 3}{\scriptsize{$\alpha$ = 3}}
\psfrag{ab = 6}{\scriptsize{$\alpha$ = 6}}
\psfrag{ab = 9}{\scriptsize{$\alpha$ = 8}}
    \includegraphics[width=0.7\columnwidth]{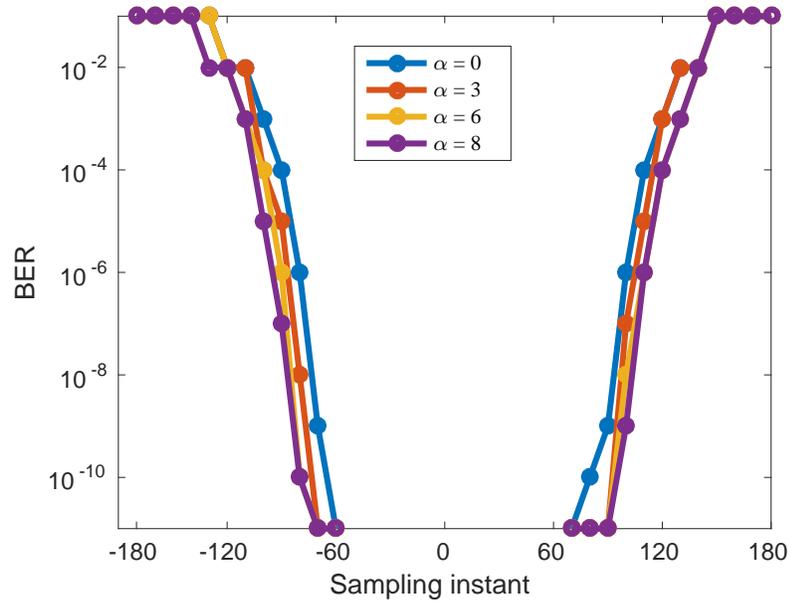}
\caption{Measured bath tub curves for a few values of the tap weight.}
\label{fig:dfe_measured_data}
\end{figure}

\begin{figure}[h!]
\centering
\includegraphics[width=0.7\columnwidth]{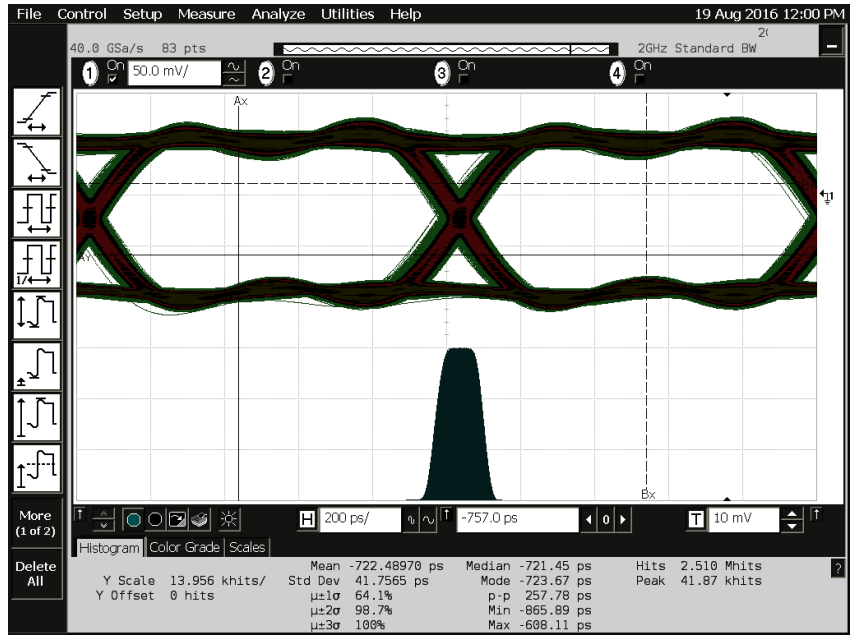}
\caption{Measured eye diagram of the recovered data.}
\label{fig:dfe_eye_measured}
\end{figure}

The circuit was tested at a frequency of 1 GHz with a supply of 1.2~V. 
The circuit consumes 145~$\mu$A of current at this frequency.
First, the DFE feedback factor was set as zero and the offset was swept
to find the code which showed the widest bath tub.
After the offset code was found the DFE tap was increased and the bath 
tubs were measured again. Fig.~\ref{fig:dfe_measured_data} shows the 
bath tub plots obtained for various values of the DFE feedback factor.
The wider bath tubs correspond to higher values of the feedback factor. 
The width of the bath tub increases by 15\% for the highest tap weight.
Fig.~\ref{fig:dfe_eye_measured} shows the eye diagram of the recovered data 
when the data is sampled at the minimum BER sampling instant.
From layout extracted simulations, the loop delay is found to be 350 ps. 

\section{Conclusions}
\label{sec:conclusion}
In this paper, we report a low latency DFE circuit with integrated 
offset compensation, built around a sense amplifier comparator with a 
switched capacitor feedback network. 
%
%
The switched capacitor circuit 
uses signals from the first stage of the sense 
amplifier comparator for selecting from precomputed bias voltages, 
thus resulting in low latency. 
The bias voltages are programmed for the sum of DFE feeback weight 
and offset to be corrected. This allows DFE and offset correction with the 
same feedback input, 
avoiding an extra offset correction input in the comparator.
A 5 bit DAC, along 
with a little logic circuitry, is used to generate the required four bias 
voltages. The circuit is designed, fabricated and tested in UMC 130 nm
CMOS technology.

\section*{Acknowledgements}
This work was supported by the Tata Consultancy Services (TCS) in the
form of student scholarships and the SMDP programme of the Government
of India in the form of CAD tool licenses.
The authors would like to thank Nagendra Krishnapura and Shanthi Pavan of 
IIT Madras, for giving access to the VLSI testing laboratory.

\bibliographystyle{elsarticle-num}
\bibliography{./ref_dfe.bib}

\end{document}